\documentclass[12pt]{article}
\usepackage{cite}
\usepackage{amsmath}
\usepackage{multicol}

\begin{document}
\title{Simple Model of Complex Reflection Behaviour in Two-Species Community}
% Use \titlerunning{Short Title} for an abbreviated version of
% your contribution title if the original one is too long
\author{Maria Yu.Senashova\footnote{Institute of Computational Modelling of RAS, 660036 Russia, Krasnoyarsk, Akademgorodok; \texttt{msen@icm.krasn.ru}}
\and Michael G.Sadovsky\footnote{Institute of Computational
Modelling of RAS, 660036 Russia, Krasnoyarsk, Akademgorodok, and
Institute of Biophysics of SD of RAS, 660036 Russia, Krasnoyarsk,
Akademgorodok; \texttt{msad@icm.krasn.ru}} \and Kristina
Kourshakova\footnote{Siberian National University, Institute of
Natural Sciences \& Humanities, 660041 Russia, Krasnoyarsk,
Svobodny pr., 79; \texttt{seriouskris@mail.ru}}}
\date{\empty}
\maketitle\vspace{3cm}

Models of spatially distributed populations and/or communities are still a matter of challenge for the
students working in population biology, ecology, environmental sciences and mathematical modelling. An
adequate model pattern to describe, model and predict the impact of spatial structure on a community
dynamics, as well as the migration processes themselves, is a key problem here. Adequate modelling of a
spatial transfer of a being is the basic difficulty here. Currently, the basic methodology addressing the
problem stands on implementation of partial differential equation of ``reaction${}\div{}$kinetics'' type.

Such approach has serious discrepancy. To be valid, the models strongly require that the beings move over
space randomly and spontaneously (aimlessly). This constraint is never met in nature; even microorganisms
control their spatial redistribution \cite{g1,g2,g3}.

Previously, there was proposed an approached to model a dynamics of a community with respect to spatial
effects based on the (micro)evolutionary principle \cite{g3,m1,m5}. In brief, the principle forces beings to
migrate in the manner improving their existence. An improvement of existence of beings is here the key
question; the answer is given by the net reproduction function \cite{m2,m3,m4,m5} $k\left(\rho,
\overrightarrow{r} \right)$. Here $\rho = \rho(\overrightarrow{r})$ is the (local) density of a population,
and $\overrightarrow{r}$ is the point in space. Obviously, the population density $\rho$ depends on the point
at space.

Net reproduction function results from two effects: the former is reproduction, and the latter is inheritance
\cite{m2,m3,m4}. If these two effects take place, then an equation of a dynamics of biological entity must
look like
\begin{equation}\label{eq:v1}
\dot{\rho}(\overrightarrow{r}, t) = \rho(\overrightarrow{r}, t)\cdot k\left(\rho(\overrightarrow{r}, t)
\right)\quad \textrm{or} \quad \rho_{t+1}(\overrightarrow{r}, t) = \rho_t(\overrightarrow{r}, t) \cdot
k\left(\rho_t(\overrightarrow{r}, t) \right)\,,
\end{equation}
for discrete time. Here $k(\rho)$ ($k(N)$, respectively) is the net reproduction function. It must be bounded
above. The equations (\ref{eq:v1}) are the equations with inheritance. A comprehensive theory of such
equations, in the most general case, including the investigation of three types of evolutionary stability see
in \cite{m5,m2,m3,m4}. The most general result is that net reproduction function $k(\rho)$ must meet the
extreme principle, for any free evolving biological community. $k(\rho)$ is maximal for those species
(entities) that sustained during the evolution; $k(\rho) = 0$ in continuous time case, and $k(\rho) = 1$ for
discrete time.

Finally, the basic issue of our model of spatially dependent dynamics of a community is that any migration
must not be a random walk, but result in a growth of net reproduction. Further, we shall consider a model in
discrete time and discrete space sites (called stations). A study of continuous model brings severe technical
problems, so we shall start from a discrete case.

\section{Model of Two-Species Community}\label{sec:1}
We shall study a dynamics of a community consisting of two species; they are supposed to be
{\sl``prey${}\div{}$predator''} related. It means, that one species exists due to external resources, but
other one lives due to the beings of the former species. It is also supposed, that both species occupy two
stations\footnote{These are the sites considered together with the environmental conditions.} and migration
means a transfer of being (of any species) from station to station. Any other movements (inevitable in real
situation) are neglected and supposed to have no effect on a community dynamics.

The dynamics of (isolated) subcommunity occupying a station is supposed to follow the discrete analogue of
classic Lotka-Volterra equation, if no migration takes place:
\begin{equation}\label{eq:sec1:1}
\begin{array}{rclcrcl}
N_{t+1}&=&N_t \cdot \left(a - bN_t - fX_t\right)&\quad&M_{t+1}&=&M_t \cdot \left(c - dM_t - gY_t\right)\\
X_{t+1}&=&X_t \cdot \left(\varepsilon fN_t - hX_t\right)&\quad& Y_{t+1}&=&Y_t\cdot \left(\varepsilon gM_t - kY_t\right)\,.\\
\end{array}
\end{equation}
Here $N_t, X_t$ are the abundances of prey and predator, respectively, at the first station, and $M_t, Y_t$
are similar variables at the second station. Parameters $a$ and $c$ determine a fertility of prey population,
in the relevant stations; parameters $b$ and $d$ describe the density-dependent self-regulation of this
population, in relevant station. Parameters $h$ and $k$ describe similar density-dependent regulation at
predator subpopulations. Parameters $f$ and $g$ describe, in general form, an efficiency of the interaction
of the beings of these two species, including a success of hunting, success of escape, etc., in corresponding
stations. Finally, $\varepsilon$ represents an efficiency of the conversion of prey biomass into the predator
biomass.

\subsection{Basic Model of Migration}\label{rekslo}
Parameter $p$, $0 \leq p \leq 1$ is a mobility of prey beings; similar, $q$, $0 \leq q \leq 1$ is mobility of
predator beings. These parameters are the transfer cost and might be interpreted as a probability of the
successful migration from one station to other; success here means that no damage for further reproduction
had taken place. Migration from station~\textbf{A} to station~\textbf{B} starts, if living conditions
``there'' are better, than ``here'', with respect to the transfer cost:
\begin{equation}\label{eq:sec1:2}
\begin{array}{rcl}
\left(a - bN_t - fX_t\right)& < &p\cdot \left(c - dM_t - gY_t\right)\,,\\ \left(\varepsilon fN_t -
hX_t\right)& < &q\cdot \left(\varepsilon gM_t - kY_t\right)\,,
\end{array}
\end{equation}
for prey and predator beings, respectively. It should be stressed, that the migration act is executed
independently by each being, while the model considers it as a population event. The backward migration
conditions are defined similarly:
\begin{equation}\label{eq:sec1:2_2}
\begin{array}{rcl}
p \cdot \left (a - bN_t - fX_t\right)& > &\left(c - dM_t - gY_t\right)\,,\\ q\cdot \left(\varepsilon fN_t -
hX_t\right)& >& \left(\varepsilon gM_t - kY_t\right)\,.
\end{array}
\end{equation}
Migration act runs each time moment $t$, for both species independently. If neither of the inequalities
(\ref{eq:sec1:2}, \ref{eq:sec1:2_2}) are fulfilled, then no migration takes place, at the given time moment
$t$. Prey migration flux $\Delta$ (predator migration flux $\Theta$, respectively) must equalize inequalities
(\ref{eq:sec1:2}, \ref{eq:sec1:2_2}):
\begin{subequations}\label{eq:sec1:3}
\begin{equation}\label{eq:sec1:3-1}
\begin{array}{c}
\left(a - b(N_t - \Delta) - fX_t\right) = p \cdot \left(c - d(M_t+p\Delta) - gY_t\right)\,,\\
\left(\varepsilon fN_t - h(X_t-\Theta)\right) = q \cdot \left(\varepsilon gM_t - k(Y_t+q\Theta)\right)
\end{array}
\end{equation}
\textrm{for the case (\ref{eq:sec1:2}), or}
\begin{equation}\label{eq:sec1:3-2}
\begin{array}{c}
p\cdot \left (a - b(N_t + p \Delta) - fX_t\right) = \left(c - d(M_t - \Delta) - gY_t\right)\,,\\
q \cdot \left(\varepsilon fN_t - h(X_t+ q \Theta)\right) = \left(\varepsilon gM_t - k(Y_t - \Theta)\right)
\end{array}
\end{equation}
\end{subequations}
for the case (\ref{eq:sec1:2_2}). Then, $\Delta$ ($\Theta$, respectively) is equal to
\begin{subequations}\label{eq:sec1:4}
\begin{equation}\label{eq:sec1:4-1}
\Delta = \frac{pc-a+bN-pdM+fX-pgY}{b+p^2d}\,, \quad \Theta = \frac{hX+ \varepsilon qgM - \varepsilon fN -
qkY}{h+q^2k}
\end{equation}
\textrm{for migration form station \textbf{A} to station \textbf{B}, and}
\begin{equation}\label{eq:sec1:4-2}
\Delta = \frac{pa-c+dM-pbN+gY-pfX}{d+p^2b}\,, \quad \Theta = \frac{kY+ \varepsilon qfN - \varepsilon gM -
qhX}{k+q^2h}
\end{equation}
\end{subequations}
for the backward migration.

Finally, let's outline how the basic model (\ref{eq:sec1:1}~-- \ref{eq:sec1:4}) works. For each time moment
$t$, a direction and the migration fluxes ($\Delta$ and $\Theta$, respectively) are determined. Then, the
species redistribute themselves according to the Eqs.\,(\ref{eq:sec1:4}). Then, the abundances of the next
generation $\{N_{t+1}, X_{t+1};\ M_{t+1}, Y_{t+1}\}$ are determined, according to (\ref{eq:sec1:1}), with the
relevant abundances of the current generation $\{\widetilde{N}_{t}, \widetilde{X}_{t};\ \widetilde{M}_{t},
\widetilde{Y}_{t}\}$ defined by (\ref{eq:sec1:3}). If no migration must take place at the current timer
moment $t$, the the stage with species redistribution is omitted.

\subsection{Reflexive Behaviour}\label{infkap}
Reflection in behaviour means an ability of a being to foresee and/or predict the behaviour of an opponent,
in a competitive behavioural act. An implementation of reflexive behavioural strategy by animals is a well
known. Not discussing here psychological or ethological aspects of such strategies implementation, let
concentrate on a simple model revealing the dynamic effects of them.

Basic model (\ref{eq:sec1:1}~-- \ref{eq:sec1:4}) does not exhibit any reflexive behaviour. An introduction of
that latter into the basic model may only be concerned with the spatial redistribution. In other words,
reflection of the optimal migration behaviour means that a being is able to ``foresee'' the migration
behaviour of a competitive species being. With respect to it, one may assume the following patterns of the
reflection in the behaviour of the species: ({\sl i})~preys reflect predators; ({\sl ii})~predators reflect
preys, and, finally, ({\sl iii})~both species reflect each other.

Thus, within the framework of our model, a reflection means that the species manifesting a reflection in the
behaviour, detects the migration conditions and chooses the migration flux according to the abundances of a
competing species, that would be produced due to the migration of that latter, not the current ones. In
case~({\sl i}) formula for $\Theta$ would remain the same, but the formula for $\Delta$ would change for
\begin{equation}\label{eq:sec1:5}
\Delta = \left\{
\begin{array}{l}
\displaystyle\frac{pc-a+bN-pdM+f\widetilde{X}-pg\widetilde{Y}}{b+p^2d}\, \qquad \textrm{or}\\[4mm]
\displaystyle\frac{pa-c+dM-pbN+g\widetilde{Y}-pf\widetilde{X}}{d+p^2b}\,,\\
\end{array}
\right.
\end{equation}
in dependence of the migration direction. Here $\widetilde{X}$ and $\widetilde{Y}$ are determined according
to~(\ref{eq:sec1:3}).

Reciprocally, $\widetilde{N}$ and $\widetilde{M}$ are determined according to~(\ref{eq:sec1:3}), for the
case~({\sl ii}), but migration flux $\Theta$ of predator would be determined by
\begin{equation}\label{eq:se1:6}
\Theta = \left\{
\begin{array}{l}
\displaystyle\frac{hX+ \varepsilon qg\widetilde{M} - \varepsilon f\widetilde{N} - qkY}{h+q^2k}\qquad
\textrm{or}\\[4mm] \displaystyle\frac{kY+ \varepsilon qf\widetilde{N} - \varepsilon g\widetilde{M} -
qhX}{k+q^2h}\,,\\
\end{array}
\right.
\end{equation}
in dependance of the migration direction.

Finally, if both species reciprocally reflect the behaviour of each other, then basic model should be changed
for the following one. On the first stage, both species determine the migration fluxes according to basic
model~(\ref{eq:sec1:3}, \ref{eq:sec1:4}). Then, they redefine the migration fluxes (and migration direction,
as well as the fact of migration) so, that each species changes the current abundances of the competitive
beings for those that could be produced due to a migration rule determined by the basic model. So, they
redefine the fluxes, redistribute themselves between the stations, and reproduce.

\section{Results and Discussion}\label{infem}
Main purpose of this paper is to figure out the sets of the parameters providing an evolutionary advantage to
a bearer of some (reflexive, or not) spatial distribution strategy. Evolutionary advantage here is understood
as an excess of the total abundance of some species realizing reflexive strategy, in comparison to the same
species in case of realization of regular (non-reflexive) strategy.

Table~\ref{tab:1} shows the result of simulation observed for the
following parameters sets:
\newcounter{N}
\begin{list}{\#\arabic{N}}{\topsep=0cm \itemsep=0cm \usecounter{N}}
\item\label{n1} $a=3.1$, $c=1.49999$, $b=0.00098$, $d=0.00099$, $h=0.00052$, $k=0.0005$,
\mbox{$f=g=0.00542$}, $\varepsilon = 0.099$, $p=q=0.99$. \item\label{n3} $a=2.5$, $c=1.49999$, $b=0.00098$,
$d=0.00099$, $h=k=0.0005$, \mbox{$f=g=0.0054$}, $\varepsilon = 0.099$, $p=q=0.99$. \item\label{n4} $a=c=1.5$,
$b=d=0.0001$, $h=0.0001$, $k=0.0005$, \mbox{$f=g=0.0059$}, $\varepsilon = 0.05$, $p=q=0.99$. \item\label{n5}
$a=c=1.5$, $b=d=0.0001$, $h=k=0.00001$, $f=0.0059$, $g=0.0049$, $\varepsilon = 0.05$, $p=q=0.99$.
\item\label{n6} $a=1.7$, $c=1.5$, $b=d=0.0001$, $h=k=0.00001$, \mbox{$f=g=0.005$}, $\varepsilon = 0.05$,
$p=q=0.99$. \item\label{n7} $a=2.2$, $c=2.1$, $b=d=0.0001$, $h=k=0.00001$, \mbox{$f=g=0.005$}, $\varepsilon =
0.05$, $p=q=0.99$.
\end{list}\vspace{3mm}

First of all, it should be said, that the basic model (\ref{eq:sec1:1}~-- \ref{eq:sec1:4}) exhibits a great
diversity of limit regimes. It may be a steady state (in both stations, for both species), limit cycles of
various length, and a complex irregular behaviour looking like a dynamic chaos. All these peculiar regimes
may be met in combinations, with respect to a station and/of a species. In general, a decrease of transfer
cost $p$ and $q$ yields a simplification of an observed regime. Both the basic model (\ref{eq:sec1:1}~--
\ref{eq:sec1:4}) and its versions implementing various reflexive strategies of spatial distribution yield an
expansion of the area of permissible parameter values, and the area of the phase space (i.e., abundance
figures). All these issues are very interesting, from the point of view of the study of the models of
optimally migrating communities, but they fall beyond the scope of our research.

\begin{table}[!h]
\caption{\label{tab:1}Comparison of various strategies of space
distribution. $S$~--- type of strategy: ${1}$~-- basic model,
${2}$~-- model {\sl i}, ${3}$~-- model {\sl ii}, ${4}$~-- model
{\sl iii}.}
\begin{center}
\begin{tabular}{|r|c|c|c|c|r|c|c|c|c|}
\hline $S$ & $N$ & $X$ & $M$ & $Y$ & $S$ & $N$ & $X$ & $M$ & $Y$
\\ \hline
\multicolumn{1}{|c}{}&\multicolumn{4}{c|}{Set of paramenter \# \ref{n1}}&\multicolumn{1}{|c}{}&\multicolumn{4}{c|}{Set of paramenter \# \ref{n3}}\\
\hline
$1$ & 2089 & 5 & 4539 & 6 & $1$ & 1513.5 & 1.4 & 1952.5 & 56.2\\
$2$ & 1930 & 20.5 & 2368 & 50 & $2$ & 1438.8 & 4.1 & 2244.6 & 55.6\\
$3$ & 1692.9 & 43.7 & 1849.9 & 41.7 & $3$ & 1524.5 & 0.47 & 1938.6 & 56.7\\
$4$ & 2142 & 0.1 & 5040 & 0.1 & $4$ & 1530 & 0.01 & 5047 &
0.01\\\hline
\multicolumn{1}{|c}{}&\multicolumn{4}{c|}{Set of paramenter \# \ref{n4}}&\multicolumn{1}{|c}{}&\multicolumn{4}{c|}{Set of paramenter \# \ref{n5}}\\
\hline
$1$ & 3697 & 18 & 3698 & 18 & $1$ & 3542 & 20.9 & 4073 & 14.3\\
$2$ & 4999 & 0.01 & 4999 & 0.01 & $2$ & 4997 & 0.01 & 4998 & 0.01\\
$3$ & 3873 & 6.4 & 3869 & 6.5 & $3$ & 3619 & 11.2 & 3739 & 8\\
$4$ & 4189 & 13.8 & 4190 & 13.8 & $4$ & 4325 & 12.5 & 4384 & 11.5\\\hline\multicolumn{1}{|c}{}&\multicolumn{4}{c|}{Set of paramenter \# \ref{n6}}&\multicolumn{1}{|c}{}&\multicolumn{4}{c|}{Set of paramenter \# \ref{n7}}\\
\hline
$1$ & 6994 & 0.05 & 4996 & 0.01 & $1$ & 5.4 & 0.1 & 4.4 & 0.04\\
$2$ & 5688 & 32.1 & 4122 & 11.6 & $2$ & 8406 & 80.5 & 7172 & 69.3\\
$3$ & 4664 & 19.5 & 2961 & 13.9 & $3$ & 3.6 & 0.07 & 3.6 & 0.1\\
$4$ & 5412 & 30.5 & 4086 & 18.7 & $4$ & 8404 & 80.5 & 7171 &
69.3\\\hline
\end{tabular}
\end{center}
\end{table}%\vspace{-6mm}

What we do, was a comparative study of those four models (basic
model and three versions with reflexive behaviour) from the point
of view of the evolutionary advantage. In fact, we tried various
combinations of the parameters, similar for all four models, in
order to identify the model that yields the highest total
abundance of the species.

To answer this question, we have carried out a series of simulations (computational experiments). We
calculated the abundance of each species, in each station, for four models with the same parameter set. Then,
the abundances of prey subpopulation (of predator subpopulation, in turn) were added. Table~\ref{tab:1} shows
the results of such comparison. The parameters yielding the observed regimes are shown below the Table.

It is evident, that there exist the parameters sets yielding an evolutionary advantage for various types of
space distribution strategies. Yet, we did not studied carefully the peculiarities of the limit regimes
relevant to each evolutionary advantageous situation, meanwhile, one may expect that the reflexive strategies
provide an advantage for rather regular limit regimes, while the non-reflexive strategy of space distribution
is advantageous for chaotic-like, complicated limit regimes.


\begin{thebibliography}{99.}

\bibitem{g1} Yu.L.Gurevich, N.S. Manukovsky, M.G. Sadovsky: \textit{Dynamics of chemical \& biological systems}. (Nauka plc. Novosibirsk 1989) pp 159--158.

\bibitem{g2} A.N. Gorban, M.G. Sadovsky: Biotechnology and Biotechnique. \textbf{2}(5), 34 (1987).

\bibitem{g3} M.G. Sadovsky: \textit{Mathematical modelling in biology \& chemistry. Evolutionary approach.} (Nauka plc. Novosibirsk 1991) pp 78--93.

\bibitem{m1} S.A. Motolygin, D.A. Chukov, M.G. Sadovsky: R.Journal of Gen.Biology, \textbf{60}, 450 (1999).

\bibitem{m5} M.G. Sadovsky: \textit{The simplest model of spatially distributed population with reasonable migration of organisms.} (arXiv q-bio.PE/0510004 2005)

\bibitem{m2} A.N. Gorban: \textit{Systems with inheritance: dynamics of distributions with conservation of support, natural selection and finite-dimensional asymptotics
}, (arXiv cond-mat/0405451 2005)

\bibitem{m3} A.N. Gorban, R.G. Khlebopros \textit{Darwin's Demon: the Idea of Optimality and Natural Selection}, (Nauka plc Moscow 1988).

\bibitem{m4} A.N. Gorban: \textit{Equilibrium encircling. Equations of chemical kinetics and their thermodynamic
analysis}, (Nauka plc Nobosibirsk 1984).

\end{thebibliography}
\end{document}